\documentclass[referee]{raa}            

\usepackage{graphicx,times}             

\newcommand{\su}{$u_{\rm SC}$}
\newcommand{\sv}{$v_{\rm SAGE}$}
\graphicspath{{figure_pdf/}}

\begin{document}

   \title{Testing Area of the SAGE Survey
}

   \volnopage{Vol.0 (2018) No.0, 000--000}      
   \setcounter{page}{1}          

   \author{Zheng Jie
      \inst{1,2}
   \and Zhao Gang
      \inst{1,2}
   \and Wang Wei
      \inst{1,3}
   \and Fan Zhou
      \inst{1}
   \and Tan Ke-Feng
      \inst{1}
   \and Li Chun
      \inst{1}
   \and Zuo Fang
      \inst{1}
   }

    \institute{Key Laboratory of Optical Astronomy, National Astronomical Observatories,
	Chinese Academy of Sciences, Beijing 100101, China; {\it gzhao@nao.cas.cn}\\
    \and
    School of Astronomy and Space Science, University of Chinese Academy of Sciences, Beijing 100049, China\\
    \and
    Chinese Academy of Sciences South America Center for Astronomy,
	China-Chile Joint Center for Astronomy,
	National Astronomical Observatories, Chinese Academy of Sciences, Beijing 100101, China\\
   }

   \date{Received~~2018 Apr 14; accepted~~yyyy~mmm~dd}

\abstract{
Sky survey is one of the most important motivations to improve the astrophysics development, especially when using new photometric bands. We are performing the SAGE (Stellar Abundance and Galactic Evolution) survey  with a self-designed SAGE photometric system, which is composed of eight photometric bands. The project mainly aims to study the stellar atmospheric parameters of $\sim$0.5 billion stars in the $\sim12,000$ deg$^2$ of the northern sky, which mainly focuses on the Galactic sciences, as well as some extragalactic sciences. This work introduces the detailed data reduction process of the testing field NGC\,6791, including the data reduction of single-exposure image  and stacking multi-exposure images, and properties of the final catalogue.
}

   \authorrunning{ Zheng J., Zhao G., Wang W., Fan Z., Tan K.-F., Li C., \& Zuo F. }    
   \titlerunning{Testing Area of the SAGE Survey}  

   \maketitle

\section{Introduction} \label{sec:intro}

The Stellar atmospheric parameters can be obtained through well-defined photometric systems, e.g., Str\"omgren-Crawford (SC) photometric system. But until now, there are few sky surveys or catalogues, e.g., GCS and HM on that, for which the limiting magnitudes are too shallow, around $V$=8 magnitude. Another recent Southern-sky survey SkyMapper, which is lead by the Australian National University (ANU), also is based on their own photometric system. In the Northern sky, we still lack a deeper sky surveys ($V\sim15$), dedicating to the stellar atmospheric parameters of a large sample.

Therefore, we are performing a deep SAGE (Stellar Abundance and Galactic Evolution) photometric  survey ($V=15$ in the S/N of $100\sigma$) on the northern sky. This paper introduces a test area chosen from the survey on which we merged single-epoch catalogues produced by the science data pipeline (SDP) into a master catalogue, and analyzed the data. The procedures of the SDP include the image correction, astrometric calibration, photometry, and the flux calibration.

In order to obtain the stellar atmospheric parameters {more accurately and more efficiently, we designed a new photometric system: the SAGE system, by combining the self-designed new filters with some existing photometric bands. This system consists of 8 filters: Str\"omgren-$u$, SAGE-$v$, SDSS $g$, $r$, $i$, $H\alpha_{wide}$, $H\alpha _{narrow}$, and DDO-$51$ (\su, \sv, $g$, $r$, $i$, $H\alpha_{w}$, $H\alpha_{n}$, and DDO$51$ respectively). The SAGE system will help to study the stellar atmospheric parameters, e.g., effective temperature, surface gravity and metal abundance, but also to study the Galactic structure and evolution. \cite{fan18} introduces the SAGE photometric system, the survey strategy and the scientific goals in detail.

The SAGE survey is a northern sky survey with the SAGE photometric system, with its $5\sigma$ depths expected to be $\sim$21.5\,mag in \su-band, $\sim$21.0\,mag in \sv-band, and $\sim$19.5\,mag in $g$-, $r$- and $i$-bands. In the SAGE survey, the observations have started since 2015 and we plan to finish the whole project, including observations, photometry, flux calibrations and astrometric calibrations in four or five years. A depth-uniformed photometric catalogue will be produced by the SAGE survey, which will help in scientific research on the Milky-Way and even extragalaxies.

\section{SAGE Photometric Survey and Observation} \label{sec:sagepss}

The SAGE survey will cover an area of $\sim$12,000 deg$^2$ of the northern sky with Decl. $\delta > -5 ^{\circ}$, excluding the bright Galactic disk ($|b|<10^{\circ}$) and the sky area of 12 hr $<$ R.A. $<$ 18 hr. The detailed coverage can be found in \cite{zheng18}.

Three telescopes are used in the SAGE survey. The first is the 90-inch (2.3-meter) Bok telescope of Steward Observatory, the University of Arizona (hereafter Bok), which is located at the Kitt Peak National Observatory; The second is the Nanshan One-meter Wide-field Telescope at Nanshan Station of Xinjiang Astronomical Observatory, Chinese Academy of Sciences (NOWT); the third is the Zeiss-1000 Telescope at Maidanak Astronomical Observatory (MAO), Ulugh Beg Astronomical Institute, Uzbek Academy of Sciences (UBAI) . \cite{fan18} and \cite{zheng18} introduced the telescopes and the relevant observations in the SAGE survey.

The SAGE survey started the observations in the autumn of 2015. By the end of Jan. 2018, observations of $g$, $r$ and $i$ bands on NOWT have been completed. Meanwhile, the observations of \su- and \sv-bands on Bok are $\sim 2/3$ completed and the observations of the two bands are expected to be completed in the end of 2019. Observations on MAO 1-M telescope is scheduled to start in the autumn of 2018. The up-to-date status of the observation progress can be checked on our website\footnote{http://sage.bao.ac.cn/surveyobs/obsfootprint.php}.

\section{Test Area} \label{sec:testarea}

In this paper, we discuss a testing area as a sample: NGC\,6791, an open cluster. We performed a series of dithered observations on it with Bok in the \su- and \sv-bands. The pointing and coverage of the multiple exposures are shown in Figure \ref{fig:ngccover}, The blue box shows the outline of this testing area; the 16 black crosses are the centers of each exposure while the black dotted box is the coverage of one exposure as $1.08^{\circ}\times1.03^{\circ}$. The red circle at center is the approximate position and size of NGC\,6791.

\begin{figure}[!htbp]
\begin{center}
\includegraphics[width=\textwidth]{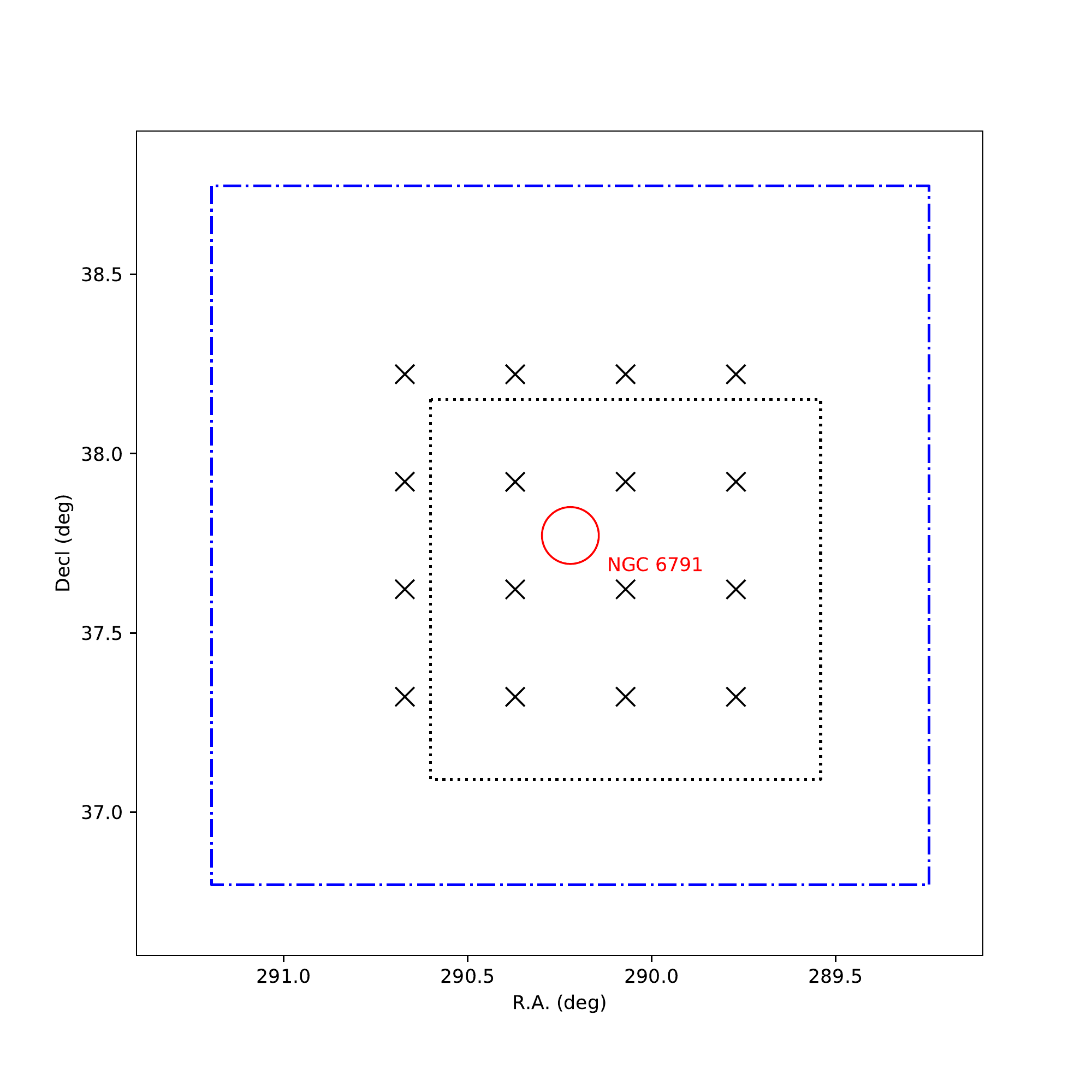}
\caption{The blue box shows the outline of this testing area; the 16 black crosses are the centers of each exposure while the black dotted box is the coverage of one exposure as $1.08^{\circ}\times1.03^{\circ}$. The red circle at center is the approximate position and size of NGC\,6791} \label{fig:ngccover}
\end{center}
\end{figure}

\section{Science Data Pipeline} \label{sec:scidata}

The science images are processed through our semi-automatic science data reduction pipeline (SDP). We stacked the  bias images and flat-fielding images for each night, and use them to correct the science images. Then we extract objects and calibrate their flux.

The Software for Calibrating AstroMetry and Photometry (SCAMP, \cite{scamp}) is used for deriving the astrometric solution, then we compute the distortion solution which is expressed by the Simple Image Polynomial (SIP, \cite{sip}). The astronomical reference catalogue adopted in our work is the Position and Proper Motion Extended (PPMX, \cite{ppmx}) and the more accurate version PPMXL Catalog of Positions and Proper Motions on the ICRS (PPMXL, \cite{ppmxl}) will be adopted in the future work.

We applied the Source Extractor (sex, \cite{sex}) to extract all the sources from images and evaluation their flux i.e. photometry. We use the Kron-like elliptical aperture photometry (MAG\_AUTO) and the corresponding uncertainties (MAGERR\_AUTO) as our main output photometry. We also perform the aperture-correction for the photometry for all the detected sources.

For the flux calibration, we convolve the flux calibrated spectrum library Hubble Space Telescope CALSPEC Flux Standards (CALSPEC, \cite{calspec}) and The Next Generation Spectral Library (NGSL, \cite{ngsl}) with the filter transmissions of the SAGE system to derive the \su{} and \sv{} magnitudes of the standard stars. We have chosen 21 standard stars from the libraries with all the spectrum types with suitable brightness for our observations. In a photometric night, we observe the standard stars in a large airmass range for dozens of times to fit the atmospheric distinction curve and the extinction coefficients, which is then applied to calibrate the flux of secondary stars that we observed. The typical uncertainty of flux calibration is $\sim$0.01\,mag.

For the \su{} and \sv{} bands, the flux calibrations are more complicated.
 For the previous version, we predicted the magnitudes from the AAVSO Photometric All-sky Survey, the 9th Data Release (APASS DS9, \cite{apass}),
 which provides the $g$, $r$, and $i$ bands photometry, by using a polynomial color-color relationship.
 The colors are derived with the photometry from
 the MILES stellar library (\cite{miles}).
 For the current version, we applied the more accurate catalogue PS1 (Panoramic Survey Telescope and Rapid Response System, Pan-STARRS, \cite{panstarrs}) to replace the APASS catalogue.

The detailed description of the data reduction steps and the results from single exposure images are discussed in \cite{zheng18}. The current SDP produces a final catalogue from the single exposure image, based on which we have done following corrections and calibrations.

\section{Reduction on Sample Data} \label{sec:sample}

In order to test the SDP, we perform a detailed and thorough reduction process on a testing area as shown in figure \ref{fig:ngccover}. Besides the data reduction for single-epoch images, we also merge the catalogues into a master catalogue, in which the photometry at the same position (R.A. and Decl.), which is regarded as one object, in different exposures will be merged.

\subsection{Balance the Flux of Images} \label{subsec:balance}

For a single-epoch image, the flux calibration is simple: just match the detected objects with reference catalogue and and calculate the zero point to calibrate the flux of the single image.

However, for the testing fields, there are 16 frames to be considered. For the first step, we need to do the internal calibration for all the images.
By matching the common stars in the overlapping part of any two images, we compute the zero-point offset and then neutralize to calibrate the two frames into the same flux level.
The balancing operation includes the following steps. First, we identify all overlapped pairs of the images in the test area, including fields overlapping by sides or corners, and multi observations of the same fields. For the test observations on NGC\,6791, any two exposures are overlapped. Then for each pair, we matched their catalogues by coordinates and compute the offset and standard deviation of the calibrated magnitudes. Now we can start the neutralization. For each image, we adjust its zero-point with the weighted mean of its relevant offsets. After adjustment of one image, we need to update the offsets related to it before adjusting other images. To prevent the transmission effect, we adjust images in a random order ,without a fixed reference image. We perform multi rounds of adjustment until the adjustments are low enough.

After the balancing, the offsets have been neutralized. As shown in Figure \ref{fig:balance}, the RMS of offsets reduces from 0.0525\,mag to 0.0055\,mag in \su-band and from 0.0244\,mag to 0.0089\,mag in \sv-band.

\begin{figure}[!htbp]
\begin{center}
\includegraphics[width=\textwidth]{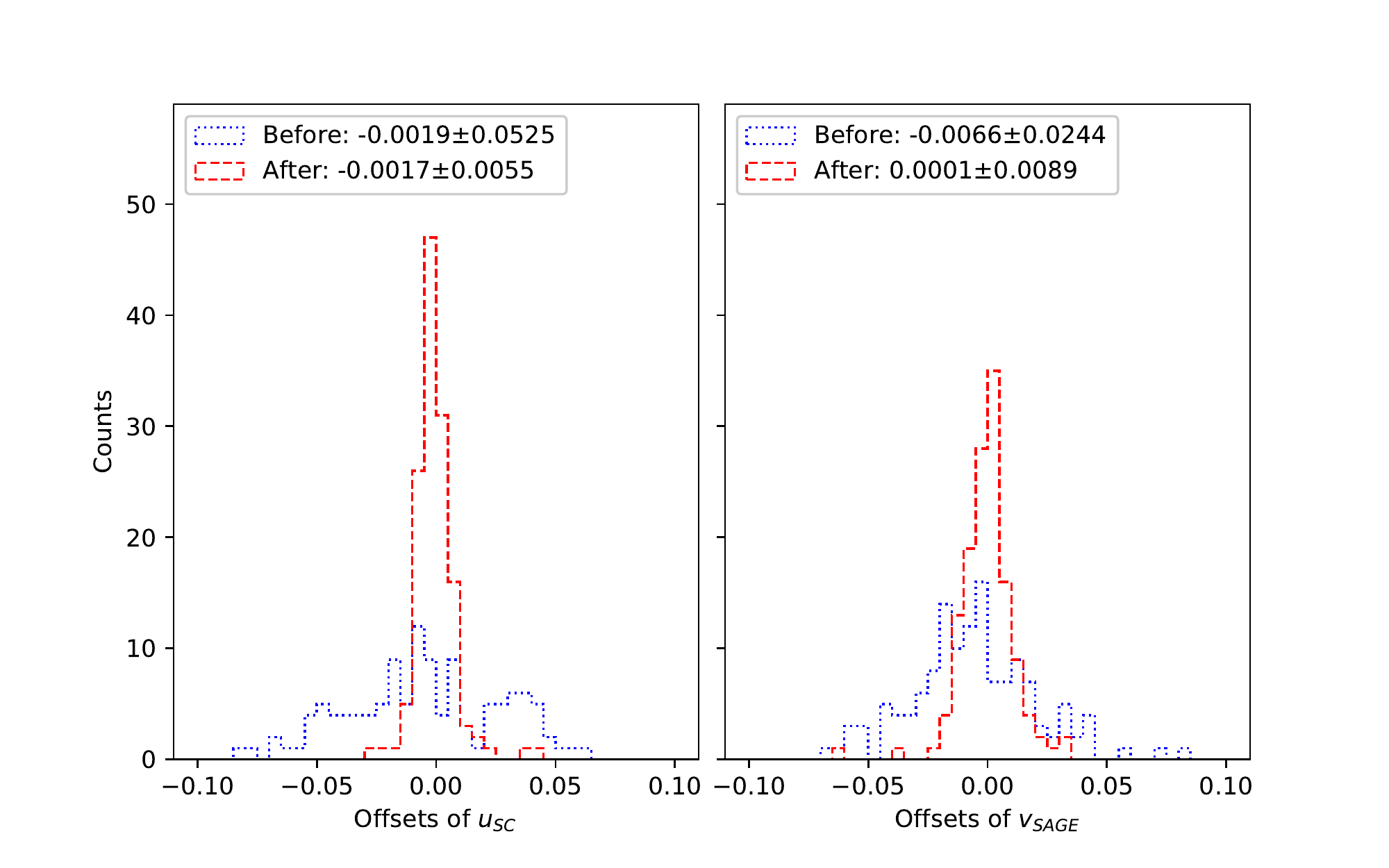}
\caption{The distribution of the zero-point offsets of overlapped images in \su-band (left) and \sv-band (right), before (blue dotted histogram) and after (red dotted histogram) the balancing operation} \label{fig:balance}
\end{center}
\end{figure}

\subsection{Columns of Catalogue} \label{subsec:fields}

After merging the catalogues of the single-exposure images, we obtained the master catalogue. Meanwhile, we also keep a total catalogue of the individual stars with single-exposure, so that we can trace the data reduction routes. The columns of the master catalogue and the total catalogue are listed in table \ref{tab:masterfields} and \ref{tab:totalfields}.

We cross-match objects by coordinates into groups from different images. The magnitudes of objects in the master catalogue are weighted means of their original magnitudes, while the errors are evaluated as the standard deviations of their original magnitudes. For those objects detected only once, we use their original magnitudes and errors. Coordinates and their errors are computed from all bands.

We use 32-bit flag columns to indicate the different situations during merging. The meaning of each bit is listed in Table \ref{tab:flags}. Flags of different photometric bands are independent. Not all bits of the flag are occupied.

We cross-match objects in different photometric bands by their positions and then cross-match with PS1 catalogue. For those object which dose not have a matched object in other bands, we put 99 as the absent magnitudes and errors, and 0 as counts.

\begin{table}[!htp]
\caption{The Columns in the Master Catalogue} \label{tab:masterfields}
\begin{center}
\begin{tabular}{ll}
\hline\noalign{\smallskip}
Parameter           & Description \\
\hline\noalign{\smallskip}
NUMBER           & Running object number \\
RA               & Right ascension of the object (J2000) \\
RA\_ERR          & Merging Right ascension of the object (J2000) \\
DEC              & Declination of the object (J2000) \\
DEC\_ERR         & Merging Declination of the object (J2000) \\
U\_COUNT         & Number of original stars contributing to this object of \su-band \\
U\_MAG\_AUTO     & Merged and calibrated Kron-like elliptical aperture magnitude of \su -band \\
U\_ERR\_AUTO     & Merging error for AUTO magnitude of \su-band \\
U\_MAG\_ISOCOR   & Merged and calibrated Corrected isophotal magnitude of \su-band \\
U\_ERR\_ISOCOR   & Merging error for corrected isophotal magnitude of \su-band \\
U\_MAG\_APERCOR  & Merged and calibrated corrected-aperture magnitude of \su-band \\
U\_ERR\_APERCOR  & Merging error vector for fixed aperture magnitude of \su-band \\
U\_MAG\_PETRO    & Merged and calibrated Petrosian-like elliptical aperture magnitude of \su-band \\
U\_ERR\_PETRO    & Merging error for PETROsian magnitude of \su-band \\
U\_FLAGS         & Merging flags of \su-band \\
V\_COUNT         & Number of original stars contributing to this object of \sv-band \\
V\_MAG\_AUTO     & Merged and calibrated Kron-like elliptical aperture magnitude of \sv-band \\
V\_ERR\_AUTO     & Merging error for AUTO magnitude of \sv-band \\
V\_MAG\_ISOCOR   & Merged and calibrated Corrected isophotal magnitude of \sv-band \\
V\_ERR\_ISOCOR   & Merging error for corrected isophotal magnitude of \sv-band \\
V\_MAG\_APERCOR  & Merged and calibrated corrected-aperture magnitude of \sv-band \\
V\_ERR\_APERCOR  & Merging error vector for fixed aperture magnitude of \sv-band \\
V\_MAG\_PETRO    & Merged and calibrated Petrosian-like elliptical aperture magnitude of \sv-band \\
V\_ERR\_PETRO    & Merging error for PETROsian magnitude of \sv-band \\
V\_FLAGS         & Merging flags of \sv-band \\
G\_MAG\_EX       & $g$ magnitude from external catalogue \\
G\_ERR\_EX       & Error of $g$-band magnitude from external catalogue \\
R\_MAG\_EX       & $r$ magnitude from external catalogue \\
R\_ERR\_EX       & Error of $r$-band magnitude from external catalogue \\
I\_MAG\_EX       & $i$ magnitude from external catalogue \\
I\_ERR\_EX       & Error of $i$-band magnitude from external catalogue \\
ID\_EX           & Object ID from external catalogue \\
\noalign{\smallskip}\hline
\end{tabular}
\end{center}
\end{table}

\begin{table}[!htp]
\caption{The Columns in the Total Catalogue} \label{tab:totalfields}
\begin{center}
\begin{tabular}{ll}
\hline\noalign{\smallskip}
Parameter           & Description \\
\hline\noalign{\smallskip}
OBS\_DATE     & Observation date, MJD \\
FILE\_NUMBER  & File number of the source \\
OBJ\_NUMBER   & Object running number from original catalogue \\
AMP           & Amplifier number where the source locating \\
X             & Object position along x inside the amplifier \\
Y             & Object position along y inside the amplifier \\
RA            & Right ascension of the object center (J2000) \\
DEC           & Declination of the object center (J2000) \\
FLAGS         & Extraction flags by SExtractor \\
FINAL\_ID     & The id number in the final catalogue \\
\noalign{\smallskip}\hline
\end{tabular}
\end{center}
\end{table}

\begin{table}[!htp]
\caption{The Meaning of Merging Flag Bits} \label{tab:flags}
\begin{center}
\begin{tabular}{ll}
\hline\noalign{\smallskip}
Bit & Description (if set to 1) \\
\hline\noalign{\smallskip}
 0    & Detected only once \\
 1    & Has other close objects but rejected while merging \\
 2    & Flux not good enough, at least 1 source is out of $3\sigma$ \\
 3    & Position not good enough, at least 1 source is out of $3\sigma$ \\
 4-27 & Reserved \\
28    & No matched object \\
29-31 & Reserved \\
\noalign{\smallskip}\hline
\end{tabular}
\end{center}
\end{table}

\subsection{The Complete Magnitudes in the Two Bands } \label{subsec:depthcount}

The testing area is only $\sim 3$ deg$^2$, which is only a tiny part compared to our survey. We show the magnitude distribution of all the detected sources in both \su-band and \sv-band from our master catalogue in Figure \ref{fig:magcount}. This turn over points are $\sim$20.0\,mag in \su-band and $\sim$21.0\,mag in \sv-band, which is in fact the complete magnitudes in the two bands.

\begin{figure}[!htbp]
\begin{center}
\includegraphics[width=\textwidth]{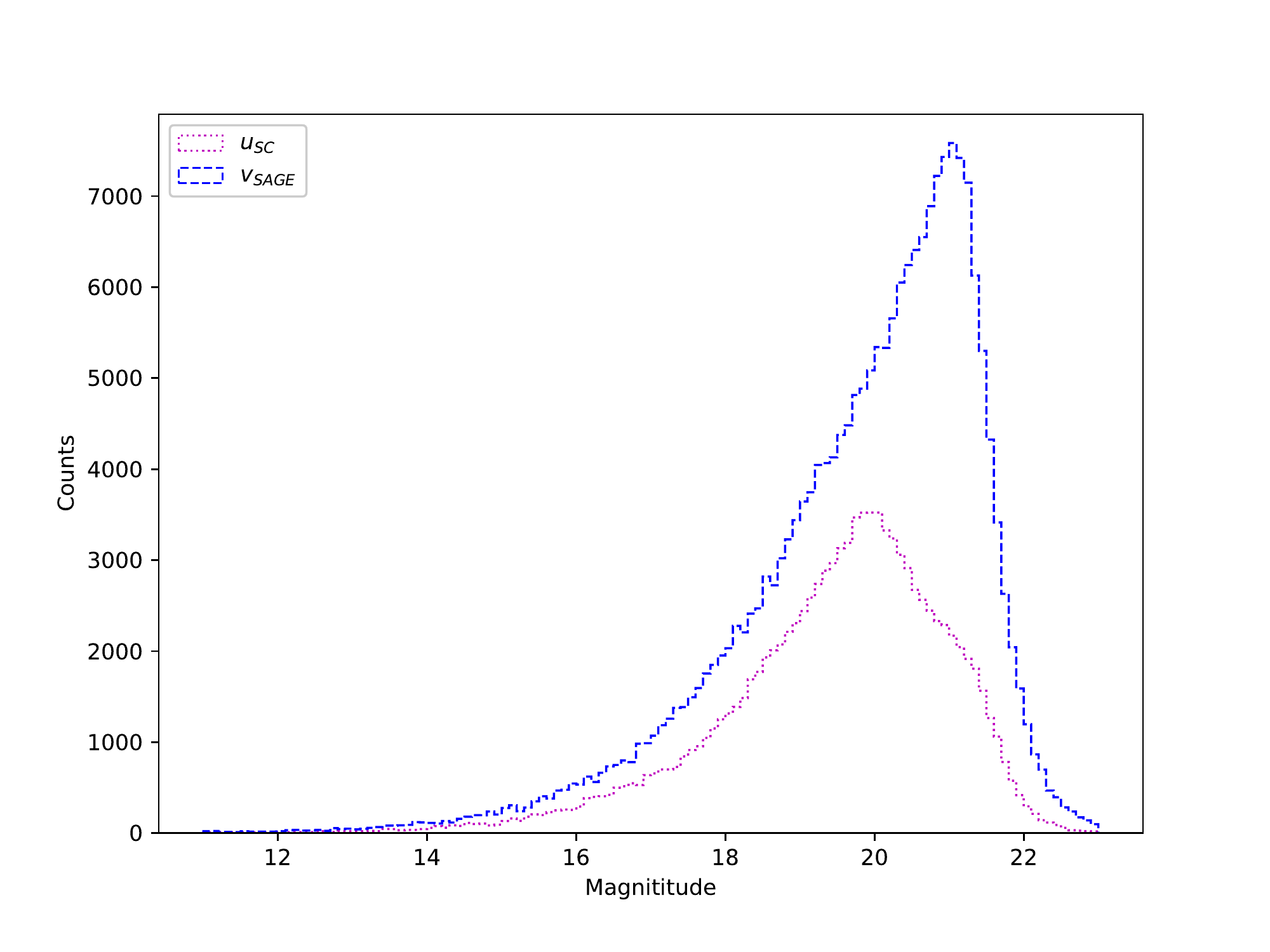}
\caption{Magnitude distributions for the two photometric bands, the \su-band complete magnitude at $\sim$20.0\,mag and the \sv-band complete magnitude at $\sim$21.0\,mag } \label{fig:magcount}
\end{center}
\end{figure}

\subsection{Color Color Diagram  of Stars} \label{subsec:color}

In order to check the zeropoints of the magnitude and colors in the flux calibration, we explore the distribution of point sources of our catalogue as well as that from stellar library in several dereddened color-color diagram in Figure \ref{fig:color2color}. We choose stars with good SNR better than 0.02 and reliable photometry.  As a comparison, we predict colors of stars from MILES by convolving the filter transmission curves with the stellar spectrum. Since we use PS1 as the reference catalogue in flux calibration, we use $g$, $r$ and $i$ bands from PS1 catalogue. The colors from MILES are overplotted as black crosses while the calibrated stars are marked with green dots.

\begin{figure}[!htbp]
\begin{center}
\includegraphics[width=\textwidth]{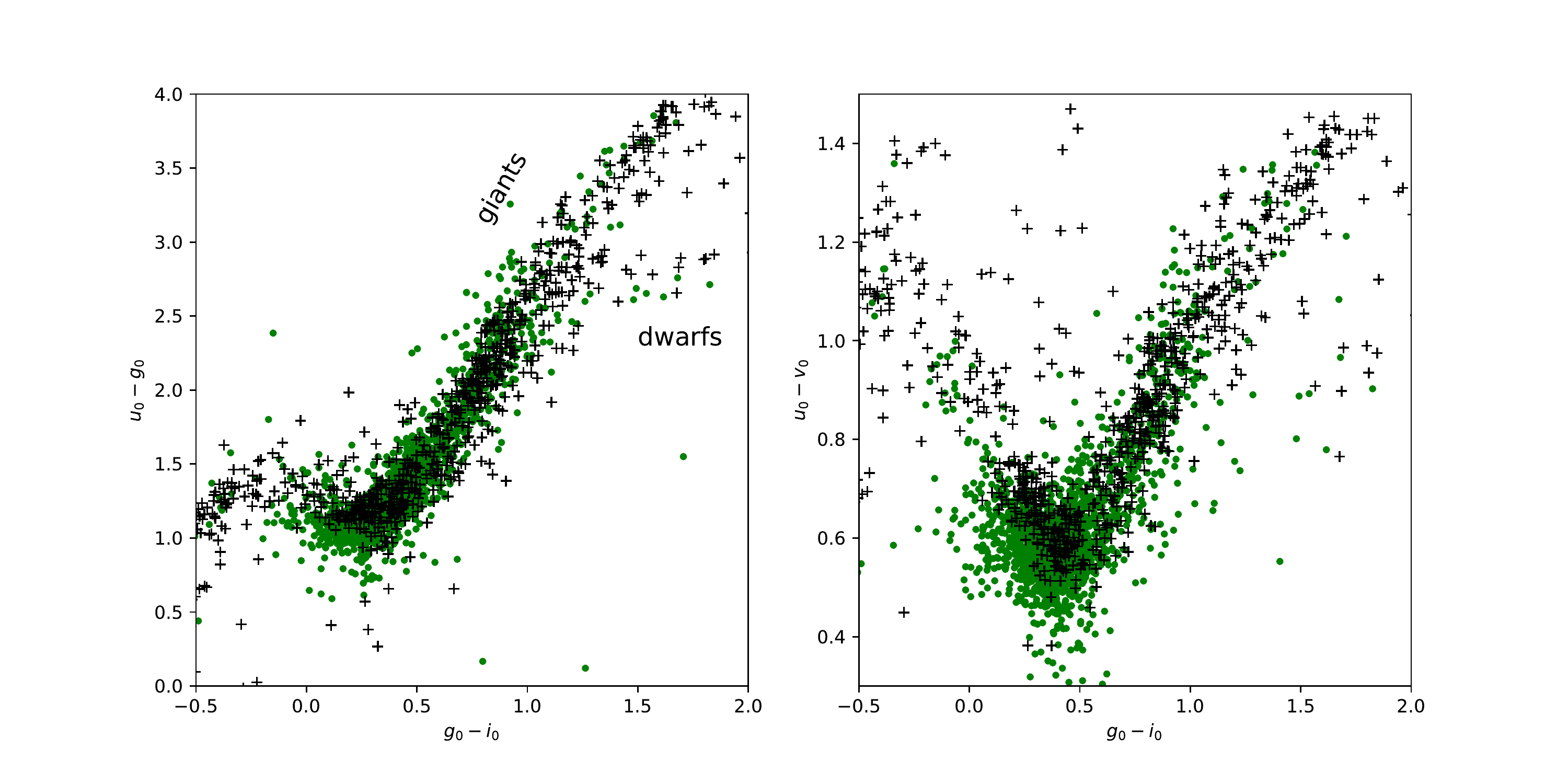}
\caption{Dereddened SAGE Colors measure for stars with good photometry (S/N higher than 50). The black crosses are prodicted colors of stars from the MILES stellar library. Clearly, we can see the  cool dwarfs branch and cool giants branch in the left panel.} \label{fig:color2color}
\end{center}
\end{figure}

\section{The Survey Future} \label{sec:future}

We are planning to complete the observation of \su- and \sv-bands in 2018, and start observations at MAO 1-m telescope from the autumn of 2018. We still need to test the whole controlling and optical system of MAO 1-m telescope.

For the data reduction , we also need to improve our pipeline. We will do photometry and calibration with more precision, and we will stack overlapped images to detect deeper objects. Furthermore, we will analyse the relationship between stellar abundance and real observed colors and try to find  a batch of the extreme metal-poor star candidates, which could be checked by the follow-up observations of other telescopes, e.g., LAMOST.

\section{Discussion} \label{sec:discuss}

We have began the SAGE survey since 2015, now we have finished about  round 2/3 in \su- and \sv-bands, all of the gri-bands of the observations. In this work we merge the catalogues from single-epoch images to a master catalogue. We cross-match objects from different images by coordinates and then neutralize their offsets to do the internal calibration. After that we have done the flux calibration with the PS1 catalogue. Furthermore we discuss the depths and colors of the test area. We are trying to analyze the zero points of the magnitude and colors of the stellar properties through these color-color diagrams. Tan et al are working on the relationship between the colors and the stellar atmospheric parameters, while Wang et al are trying to use the deep learning methods to derive the stellar atmospheric parameters with the colors by a series of training.

The SAGE photometric system is a self-designed photometric system with high sensitivity to stellar atmospheric parameters. We expect to obtain 8 colors of about 500 million stars and derive their stellar atmosphere parameters. Meanwhile, we can also retrieve reliable extinction map with  the $H\alpha_{w}$, $H\alpha_{n}$ bands photometry. The SAGE survey will become important observation resource for stellar physics and structure and evolution of the Milky-way Galaxy. At the same time, this will also provide data for the research of extragalactic objects.

\begin{acknowledgements}
Prof R. Green, Prof X. Fan, and other astronomers in the Steward Observatory, the University of Arizona, provide greatly help in observation and data reduction, thanks for them. And we also receive advices from BATC group of NAOC. We appreciate help from all of them.
 This work is supported by the National Science Foundation of China (11373003, 11673030, \& U1631102), National Key Basic Research Program of China (973 Program, 2015CB857002), and National Program on Key Research and Development Project (2016YFA0400804).
\end{acknowledgements}


\label{lastpage}

\end{document}